\documentclass[%
 reprint,
 amsmath,amssymb,
 aps,
]{revtex4-1}

\usepackage{graphicx}
\usepackage[caption=false]{subfig}

\usepackage{enumerate}

\usepackage{dcolumn}
\usepackage{bm}

\begin{document}

\preprint{APS/123-QED}

\title{Propagation of Surface Plasmon Polaritons in Thin Films of Topological Insulators}

\author{Yury Deshko}
 \affiliation{City College of New York, CUNY}
 
\author{Lia Krusin-Elbaum}
\affiliation{City College of New York, CUNY}%

\author{Vinod Menon}%
\affiliation{City College of New York, CUNY}%


\author{Alexander Khanikaev}
\affiliation{Queens College, CUNY}%
\author{Jacob Trevino}
\affiliation{Advanced Science Research Center, CUNY}%


\date{\today}

\begin{abstract}
The propagation of surface plasmon polaritons in thin films of topological insulators is studied. The materials considered are second generation three dimensional topological insulators Bi$_2$Se$_3$, Bi$_2$Te$_3$, and Sb$_2$Te$_3$. Dispersion relations and propagation lengths are estimated numerically, taking into account the variation of bulk dielectric functions of topological insulators as well as substrate using the Drude-Lorentz model. Key factors affecting propagation length are identified and ways to modify the dispersion relations are suggested. The explanation of the apparent discrepancy between the experimental data for Bi$_2$Se$_3$ and theory is proposed.
\begin{description}
\item[PACS numbers] 78.67.-n, 73.20.Mf
\end{description}
\end{abstract}

\maketitle


\def \sfw {1.3\textwidth}
\def \grafw {0.45\textwidth}

\section{\label{sec:intro}Introduction}

Since their discovery, three dimensional topological insulators (TIs)\cite{AndoReview, HasanKaneReview, QiSCZReview} have attracted enormous interest, both in theory \cite{FuKaneMele1, TI3DZhang, Yazyev} and in experiment \cite{Chen178, Hsieh, Wang}, owing to the unconventional character of gapless topological surface states hosting ``massless'' helical electron liquid \cite{Raghu}. These surfaces are envisioned as a potentially disruptive platform for a wide range of frontier technological applications, from  spintronics \cite{Appelbaum} and fault-tolerant quantum computing based on Majorana fermions \cite{Kitaev}, to terahertz optics and plasmonics \cite{YiPing}.

Here we numerically study the propagation of surface plasmon polaritons (SPPs) -- coupled oscillations of surface charges and electromagnetic field  \cite{Maier} -- in nano-meter thin films of topological insulators Bi$_2$Se$_3$, Bi$_2$Te$_3$, and Sb$_2$Te$_3$. Motivated by recent experimental observation of SPPs in Bi$_2$Se$_3$ in the far-infrared range \cite{DiPietro}, we resolve several outstanding issues. (i) We find the dispersion relations and propagation lengths of SPPs in Bi$_2$Se$_3$, Bi$_2$Te$_3$, and Sb$_2$Te$_3$ in the far-IR, using realistic material parameters allowing the comparison of these materials' potential in plasmonics; (ii) identify key parameters determining the propagation length and demonstrate that the latter can be enhanced by two orders of magnitude in some cases -- the finding which can be crucial for practical applications; (iii) revisit the problem advanced by Stauber et al.\cite{Stauber} regarding the proper analysis of the experimental data in Ref.~\cite{DiPietro} and propose a simple solution to it. (iv) analyze the effect of stacking of TI films and dielectric into 1D superlattice with the goal to modify the dispersion relation of SPPs.

Since optical response of materials depends on their bulk dielectric function $\epsilon(\omega)$, it is imperative to account for its possible variation. To properly address the questions (i-iv) we approximate $\epsilon(\omega)$ using the Drude-Lorentz model combined with available experimental data on far-IR optical properties of the materials under investigation. This methodology is the main and essential difference between our approach and the usual take on SPPs in TI films.

\section{\label{sec:background}Background}

\begin{figure}[h!]
	\centering
	\includegraphics[scale=0.5]{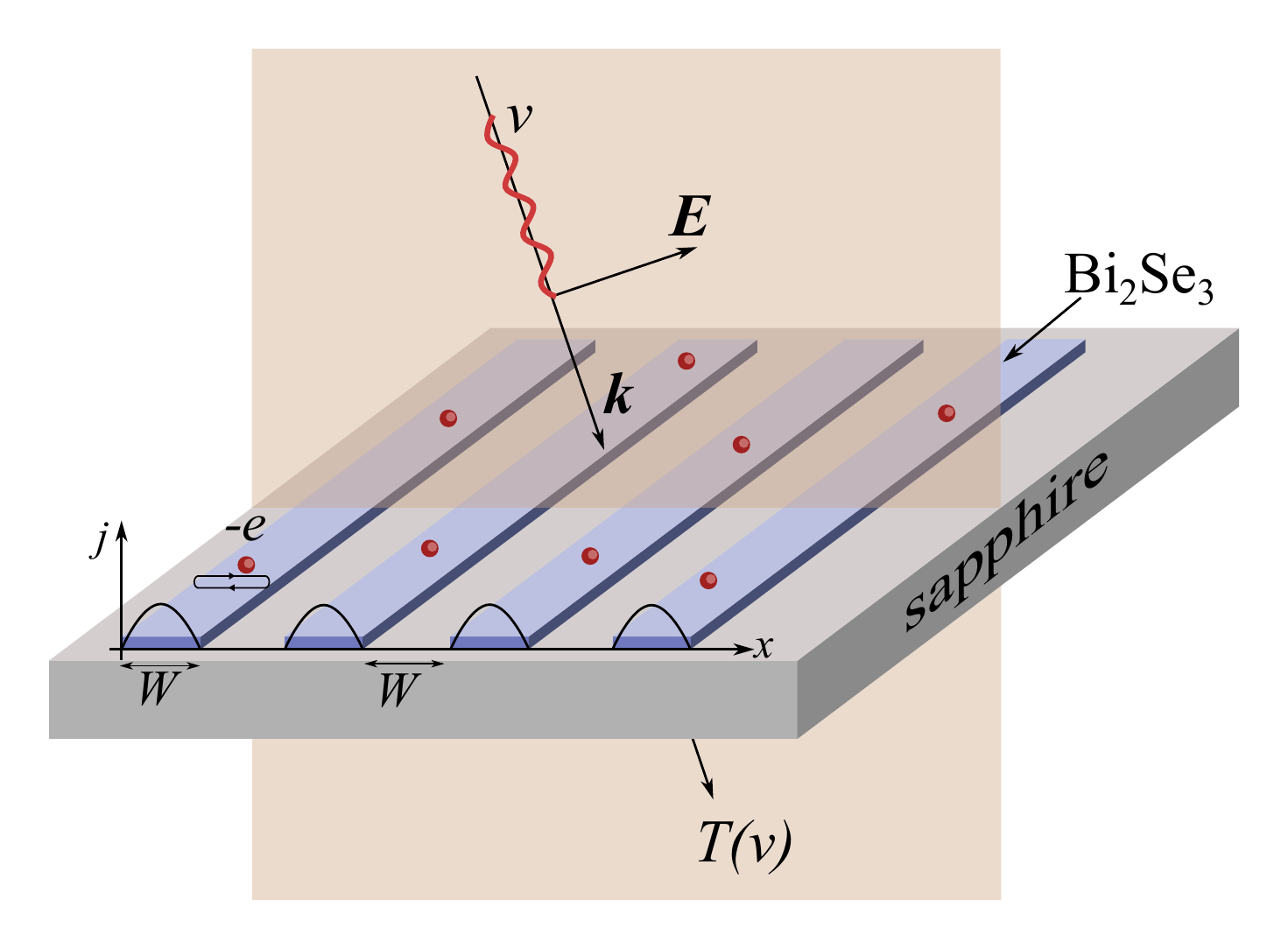}
	\caption{Schematics of the experiment for observing SPP resonances in Bi$_2$Se$_3$ (see Ref.~\cite{DiPietro}). Thin film, MBE grown on sapphire, is etched into a periodic array of ribbons of the width $W$, separated by a gap of the same width. Incident linearly polarized  far-IR light excites standing waves of collective oscillations of surface charges. The resonant excitation for given ribbon width $W$ is observed in the drop of the transmitted light intensity $T(\nu)$. The surface current density $j$, plotted on the vertical axis, describes standing waves with the period $2W$. }
	\label{fig:diPietroExperiment}
\end{figure}

Experimentally SPPs in TIs were first studied  by Di Pietro et al. \cite{DiPietro} in ribbons of Bi$_2$Se$_3$, where standing SPP waves with resonant frequencies defined by the ribbons width were formed. The idea of the experiment is illustrated in Fig.~\ref{fig:diPietroExperiment}. As the size and spacing $W$ of ribbons was varied the spectral position of SPP resonances shifted, allowing to map the dispersion curve $\omega (q)$ for the structures. The correspondence between the experiment and theoretical prediction, $\omega (q)\propto \sqrt{q}$, based on the Dirac fermions description (see e.g. Ref.~\cite{DasSarma1}) was remarkable. However, this result was later critically analyzed by Stauber et al. \cite{Stauber}, who noted that when the long range Coulomb interaction between top and bottom surfaces is properly taken into account, the experiment should have shown the sensitivity to the thickness of the films as well as to the bulk dielectric function of Bi$_2$Se$_3$. 
 
More accurate analytical expression for $\omega(q)$ was derived in Ref.~\cite{Stauber} in the long-wavelength limit ($qd\ll 1$):
\begin{equation}
	\omega^2 = \frac{v_F k_F e^2}{\varepsilon_0 h}\cdot\frac{q}{\epsilon_T+\epsilon_B+qd\epsilon_{TI}}
	\label{eq:SPPDispersion1}
\end{equation}
Here $\epsilon_{T}$ and $\epsilon_{B}$ are bulk dielectric functions of the media on top and bottom of the TI film. When $qd\epsilon_{TI}\ll (\epsilon_T+\epsilon_B)$ the well-known dispersion relation for two dimensional plasmons is recovered with $\omega\propto \sqrt{q}$. This simplified dispersion relation was used to fit the experimental results in Ref.~\cite{DiPietro}. However, Stauber et al. pointed out that for $\epsilon_{T} = 1$ (air), $\epsilon_{B} = 10$ (sapphire), and $\epsilon_{TI} = 100$  the term $qd\epsilon_{TI}$ in Eq.~(\ref{eq:SPPDispersion1}) can not be neglected. Indeed, the requirement $qd\epsilon_{TI}\ll (\epsilon_T+\epsilon_B)$ leads to $qd \ll 0.1$, which is not satisfied for the majority of data points in Ref.~\cite{DiPietro}: $qd$ = 0.02, 0.05, 0.05, 0.08, 0.2 for $W$ = 2.0, 2.5, 4.0, 8.0 and 20 microns, respectively. In order to satisfactorily fit the experiment with a more appropriate dispersion relation (\ref{eq:SPPDispersion1}), Ref.~\cite{Stauber} assumed additional contribution to the optical response from two dimensional spin-degenerate electron gas close to the surface. Although not excluding this possibility, we demonstrate that such an assumption is not necessary. Instead we reconcile the experiment \cite{DiPietro} with the theory \cite{Stauber} by calculating the dispersion relation without the restriction $\epsilon=\textrm{ const}$ for both substrate and the TI bulk.

For sapphire, which is often used as a growth substrate, the bulk dielectric function in far-IR (below 400 cm$^{-1}$) can be approximated by
\begin{equation}
	\epsilon (\omega, \textrm{cm}^{-1}) = n_0^2 + (n_0^2 - 1)(\lambda \omega)^2 + i\gamma (n_0^2 - 1)(\lambda \omega),
	\label{eq:epsSapphire}	
\end{equation}
where $n_0 = 3.2$, $\lambda = 20.4\times 10^{-4}$ cm, and $\gamma = 0.036$ are experimentally determined parameters \cite{RobertsCoon}. Although sapphire is optically anisotropic, this expression gives the refractive index for ordinary and extraordinary rays with $10$\% accuracy. Above 400 cm$^{-1}$ bulk dielectric function of sapphire begins to change drastically due to the presence of IR-active modes of lattice vibrations \cite{Schubert} and then more careful modeling is required.

The bulk optical properties of Bi$_2$Se$_3$, Bi$_2$Te$_3$, and Sb$_2$Te$_3$ in the far and mid-IR are also relatively well known from reflectance measurements \cite{Wolf, Richter}. The analysis of experimental data suggests that isotropic Drude-Lorentz model with 3 or 4 oscillators can quite satisfactorily describe the overall features of reflectance spectra in the range from 50 cm$^{-1}$ to 1000 cm$^{-1}$ for all three materials \cite{Wolf}. The bulk dielectric function in the far-IR range of interest can therefore be approximated with the expression
\begin{equation}
	\epsilon (\omega, \textrm{cm}^{-1}) = \epsilon_\infty - \frac{\omega_D^2}{\omega^2 + i\omega\gamma_D} + \sum\limits_{j = 1}^{j = 3,4} \frac{\omega_{pj}^2}{\omega_{0j}^2 - \omega^2 - i\omega\gamma_{j}}.
	\label{eq:DrudeLorentz}
\end{equation}

\begin{figure}[h]
	\centering
	\includegraphics[scale=0.5]{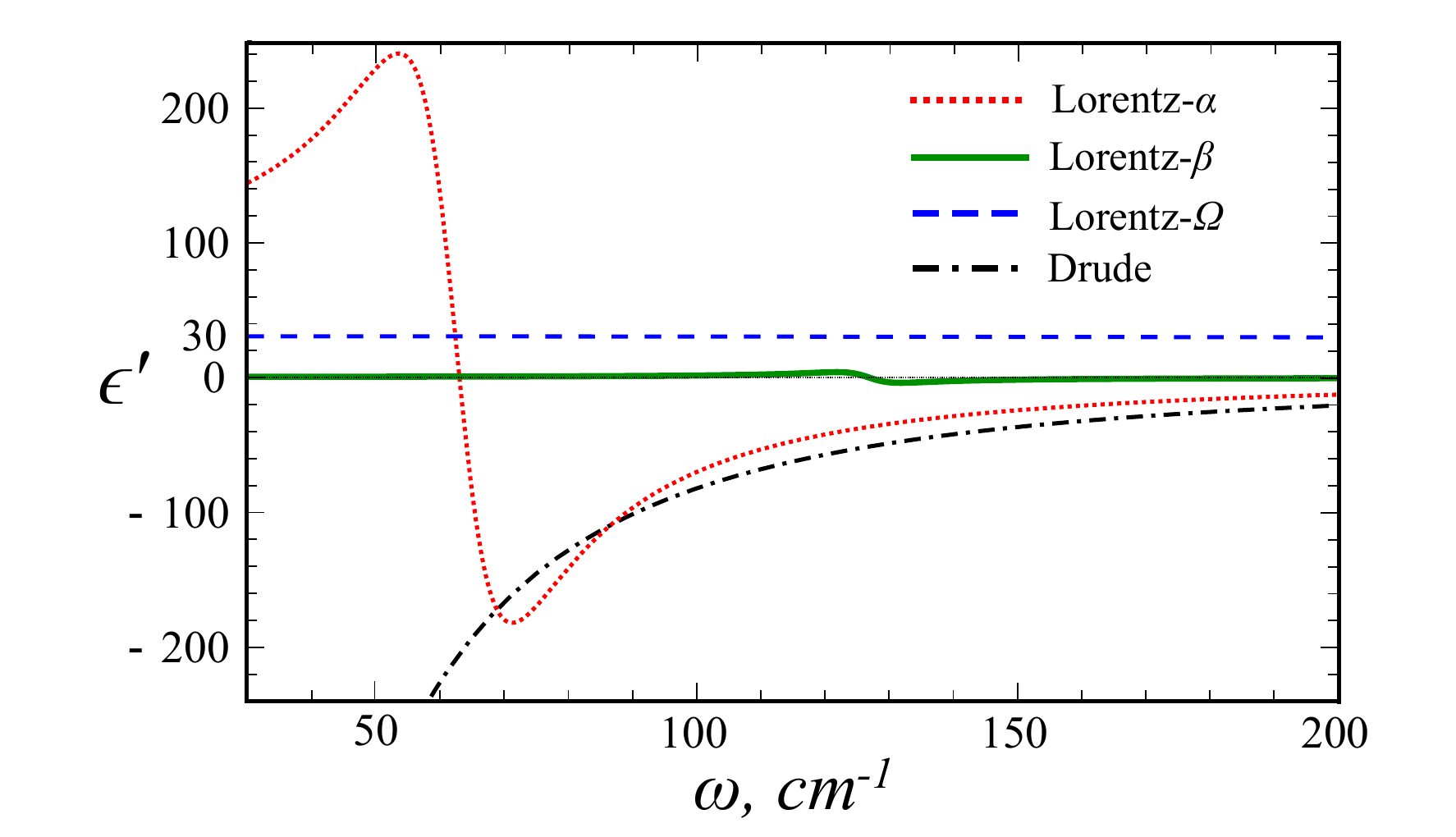}
	\caption{Contributions of various terms of the Drude-Lorentz model into the real part of the far-IR bulk dielectric function of Bi$_2$Se$_3$. Two IR-active phonon modes (Lorentz-$\alpha$ with $\omega \approx 61$ cm$^{-1}$ and Lorentz-$\beta$ with $\omega \approx 133$ cm$^{-1}$) correspond to the first two terms in the Drude-Lorentz model (\ref{eq:DrudeLorentz}). The oscillator Lorentz-$\Omega$ accounts for higher frequency absorption, in this case at $\omega \approx 2029.5$ cm$^{-1}$ = 252 meV which is close to the band-gap energy of Bi$_2$Se$_3$ bulk \cite{Richter}.}
	\label{fig:DrudeTerms}
\end{figure}
\begin{figure}[h]
	\centering
	\includegraphics[scale=0.6]{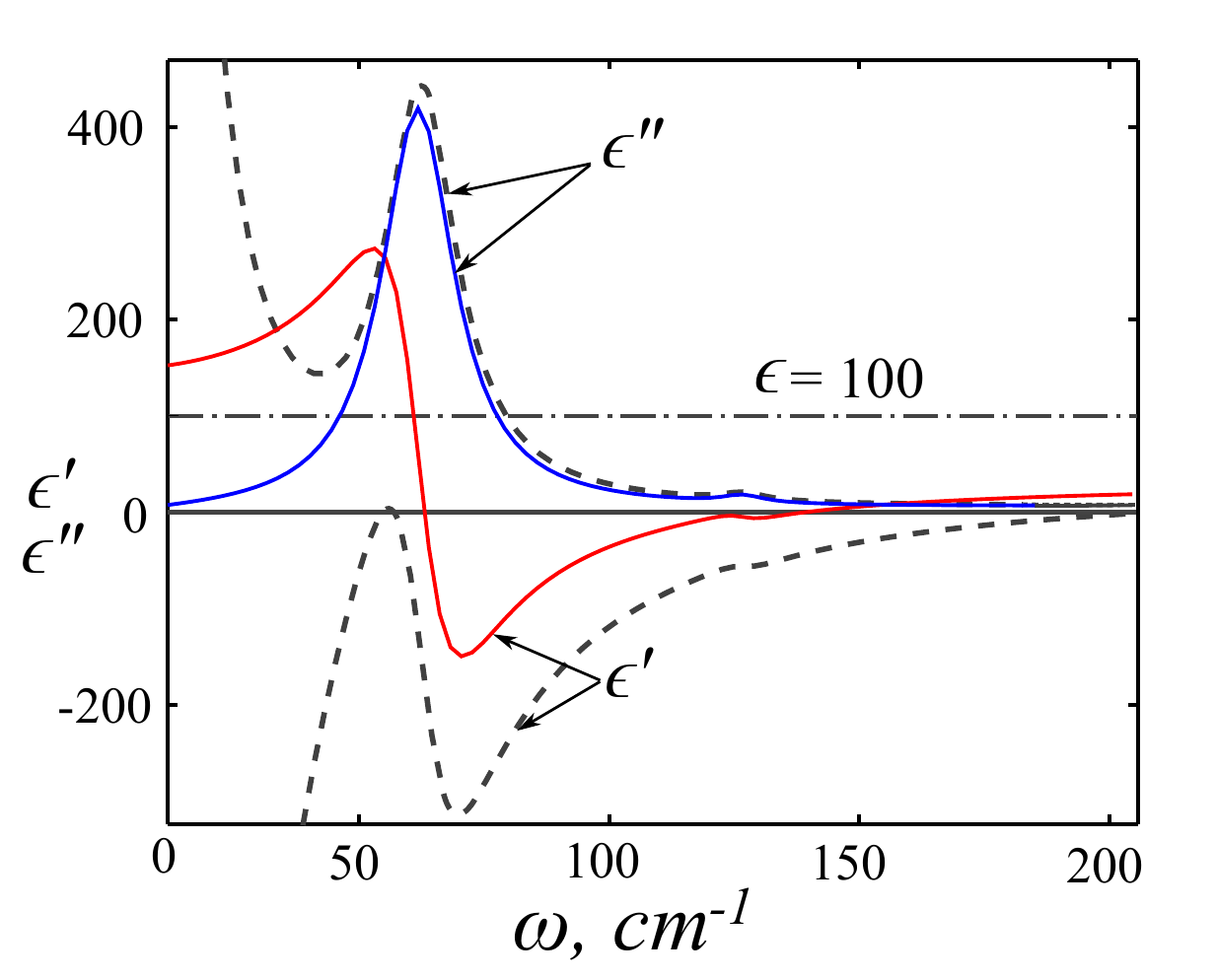}
	\caption{Real $\epsilon'$ and imaginary $\epsilon''$ parts of the bulk dielectric function of Bi$_2$Se$_3$, according to the Drude-Lorentz model. Solid lines show the bulk dielectric function without the Drude term, e.g. in a perfect crystal at low temperature. Dashed lines reproduce the bulk dielectric function with the Drude term present. The effect of Drude term is very strong at these frequencies.}
	\label{fig:LWolfE1E2}
\end{figure}

\begin{table}
\begin{center}
	\item[]
    \begin{tabular}{@{} llll}
    \hline\hline    
    Oscillator & $\omega_0$, cm$^{-1}$ & $\omega_p$, cm$^{-1}$ & $\gamma$, cm$^{-1}$ \\ \hline
    Drude & 0 & 908.66 & 7.43 \\
    Lorentz, $\alpha$ & 63.03 & 675.9 & 17.5 \\
    Lorentz, $\beta$ & 126.94 & 100 & 10 \\
    \hline
	Lorentz, $\Omega_{gap}$ & 2029.5 & 11249 & 3920.5 \\
		\hline\hline
    \end{tabular}	
	\caption{Parameters for the Drude-Lorentz model with $\epsilon_\infty = 1$, extracted from the reflectance measurements on Bi$_2$Se$_3$ at room temperatures in far and mid-IR \cite{Wolf}.}
	\label{tbl:DLWolf}
\end{center}
\end{table}

\begin{table}
\begin{center}
    \begin{tabular}{@{} llll}
    \hline\hline
    Oscillator & $\omega_0$, cm$^{-1}$ & $\omega_p$, cm$^{-1}$ & $\gamma$, cm$^{-1}$ \\ \hline
    Drude & 0 & 933.3 & 23.3 \\
    Lorentz, $\alpha$ & 63.3 & 566.7 & 10.0 \\
    Lorentz, $\beta$ & 133.3 & 66.7 & 3.33 \\
	Lorentz, $\Omega_{gap}$ & free & free & free \\
	\hline\hline
    \end{tabular}	
	\caption{Parameters for the Drude-Lorentz model for the unpatterned Bi$_2$Se$_3$ thin films at 300 K, see Supplementary Material to Ref.~\cite{DiPietro}.}
	\label{tbl:DLDiPietro}
\end{center}
\end{table}

Parameters for Bi$_2$Se$_3$ at room temperature \cite{Wolf} are shown in Table \ref{tbl:DLWolf}. Table \ref{tbl:DLDiPietro} shows similar numbers for thin films of Bi$_2$Se$_3$ used in Ref.~\cite{DiPietro}. Fig.~\ref{fig:DrudeTerms} shows the contribution of each term in the Drude-Lorentz model into the bulk $\epsilon(\omega)$ for this material, while Fig.~\ref{fig:LWolfE1E2} demonstrates that $\epsilon(\omega)$ varies significantly over the spectral range 10 cm$^{-1}$  to 200 cm$^{-1}$. We note that the contribution to $\epsilon(\omega)$ from free electrons in the bulk, represented by the Drude term in Eq.~(\ref{eq:DrudeLorentz}), is substantial. It may be argued that at low frequencies the optical response is dominated by the surface states and the bulk contribution is small \cite{Aguilar}, but if the exact contribution of the bulk Drude term is not known it must be set as a free parameter. Indeed, considering that 1) Bi$_2$Se$_3$, Bi$_2$Te$_3$, and Sb$_2$Te$_3$ all have relatively narrow band gaps, 2) the Fermi level of many thin films lies in the bulk conduction or valence band, and 3) intrinsic bulk defects increase the free carrier concentration, it is important to fully consider the bulk Drude term.

According to Ref.~\cite{Wolf}, the Drude-Lorentz model for Bi$_2$Te$_3$ at room temperature requires the Drude term with $\omega_p = 5651.5$ cm$^{-1}$, $\gamma = 111.86$ cm$^{-1}$ and a single Lorentz oscillator with $\omega_0 = 8386.6$ cm$^{-1}$, $\omega_p = 66024$ cm$^{-1}$, $\gamma = 10260$ cm$^{-1}$ ($\epsilon_\infty = 1$). Sb$_2$Te$_3$ needs only the Drude term with $\omega_p = 6906.7$ cm$^{-1}$, $\gamma = 183.69$ cm$^{-1}$ ($\epsilon_\infty = 51$). These numbers, compared to the data for Bi$_2$Se$_3$, indicate that at room temperature both Bi$_2$Te$_3$ and Sb$_2$Te$_3$ exhibit highly metallic behavior in their optical response. This is in correspondence with the fact that the bulk band gap of Bi$_2$Te$_3$ and Sb$_2$Te$_3$ is smaller than the band gap of Bi$_2$Se$_3$ \cite{TI3DZhang}. Room temperature optical response of Bi$_2$Te$_3$ and Sb$_2$Te$_3$ therefore seems to be dominated by free electrons in the bulk.

Low temperature reflectance measurements on Bi$_2$Te$_3$ and Sb$_2$Te$_3$ in the far-IR reveal the presence of IR active vibrational modes with frequencies in the range 20 cm$^{-1}$ to 200 cm$^{-1}$. For Bi$_2$Te$_3$ see for example Fig.~6 in Ref.~\cite{Richter}. Such modes are characteristic to the whole family of rhombohedral  $V_2-VI_3$ compounds and are represented by the oscillators ``Lorentz $\alpha$'' and ``Lorentz $\beta$'' in the Drude-Lorentz model. 

We emphasize the importance of taking these modes into account in order to adequately estimate the bulk $\epsilon(\omega)$ of Bi$_2$Se$_3$, Bi$_2$Te$_3$, and Sb$_2$Te$_3$ in the far-IR. By doing so we are able to determine more realistic dispersion relations for SPPs and their propagation lengths in these materials, which is essential for potential applications.

\section{\label{sec:method}Method}

We now can calculate dispersion relations and estimate propagation lengths for SPPs in thin TI films, following the approach of Ref.~\cite{Stauber}. A film of thickness $d$ supports two SPP modes (see Refs.~\cite{McDonald, DasSarma1}) and in the following we will focus on the higher-frequency optical mode. In the general case the dispersion relation for this mode can be found by solving the non-linear equation $\displaystyle{\textrm{det}\lbrack 1 - v(q,d)\chi_0(\omega, q)\rbrack = 0}$, where the matrix $v(q,d)$ incorporates intra- and interlayer Coulomb interactions between the top and bottom surface of TI film, and $\chi_0(\omega, q)$ is charge density and transverse spin susceptibility tensor (see Eqs. (2)-(7) in Ref.~\cite{Stauber}). 

In many cases of interest the criterion for the long-wavelength limit, $qd\ll 1$, is satisfied and the dispersion relation (\ref{eq:SPPDispersion1}) can be used. Indeed, consider how the SPPs are excited using the etched grating made of TI ribbons with the period $2W$ \cite{DiPietro}. The wave-vector imposed by the grating is given by $q = 2\pi/(2W)$ and the criterion $qd\ll 1$ leads to the requirement $W\gg \pi d$. For TI films with thicknesses $d = 10-100$ nm the necessary limit is satisfied if $W \gg 30-300$ nm. For the gratings with the widths of several microns one can safely use the Eq. (\ref{eq:SPPDispersion1}).

To account for the variation of the bulk dielectric functions $\epsilon_{TI}(\omega)$,  $\epsilon_{B}(\omega)$, and $\epsilon_{T}(\omega)$, it is convenient to invert Eq.~(\ref{eq:SPPDispersion1}):
\begin{equation}
	q(\omega) = \frac{A\omega^2 \lbrack \epsilon_{T}(\omega)+\epsilon_{B}(\omega) \rbrack}{1-A\omega^2 d \epsilon_{TI}(\omega)},\quad A \equiv \frac{\varepsilon_0 h}{v_F k_F e^2}.
\end{equation}

Substituting the expressions for $\epsilon_{B}$ and $\epsilon_{TI}$ with appropriate material parameters (Eqs.~(\ref{eq:epsSapphire}) and (\ref{eq:DrudeLorentz}), respectively), one can find the wave-vector $q(\omega) = q'(\omega) + i q''(\omega)$ for any frequency. The real part $q'(\omega)$ yields the dispersion relations $\omega(q')$, and the imaginary part $q''(\omega)$ determines the effective propagation length $l \equiv 1/( 2q''(\omega) )$ -- the parameter important for applications of propagating SPPs.

In addition to bulk dielectric functions $\epsilon_{T}$, $\epsilon_{TI}$, and $\epsilon_{B}$, the results of calculations depend on film thickness, the Fermi velocity, and the Fermi level. Below we present calculations for the representative values of these parameters. The Fermi velocity is set to $v_F = (5\pm 1)\times 10^5$ m/s \cite{TI3DZhang, OrbitalTexture}. The thicknesses $d$ = 15 nm, 30 nm, 60 nm, 120 nm and the Fermi energies 50 meV and 500 meV will be considered, corresponding to the Fermi wave-vectors  $k_F = 0.15\times 10^9$ m$^{-1}$ and $1.52\times 10^9$ m$^{-1}$, respectively. Due to monotonic behavior of the dispersion curves and propagation lengths we omit the intermediate values of the Fermi levels. The range of frequencies 30 cm$^{-1}$ to 200 cm$^{-1}$ is chosen in order to relate our numerical results to the available experimental data on SPP dispersion relation in Bi$_2$Se$_3$ \cite{DiPietro}.

\section{\label{sec:resdis}Results and Discussion}

\subsection{Bi$_2$Se$_3$} 

\begin{figure}[h!]
    \centering
		\subfloat[Subfigure 1 list of figures text][$k_F = 0.15\times 10^{9}$  m$^{-1}$, $E_F = 50$ meV.]{
		\includegraphics[scale=0.35]{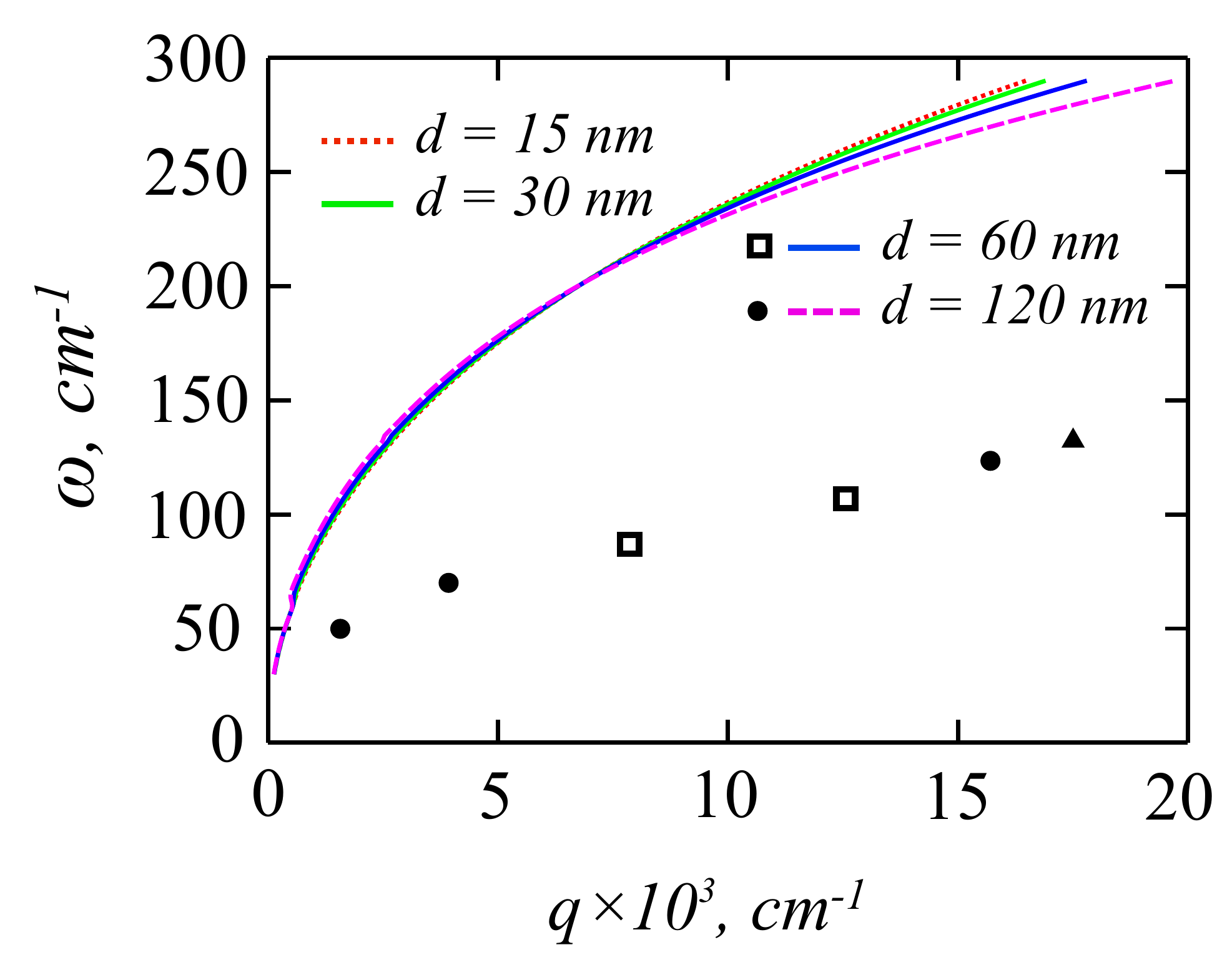}
		\label{fig:subfig1}}
		\qquad	
		\subfloat[Subfigure 4 list of figures text][$k_F = 1.52\times 10^{9}$ m$^{-1}$, $E_F = 500$ meV.]{
		\includegraphics[scale=0.35]{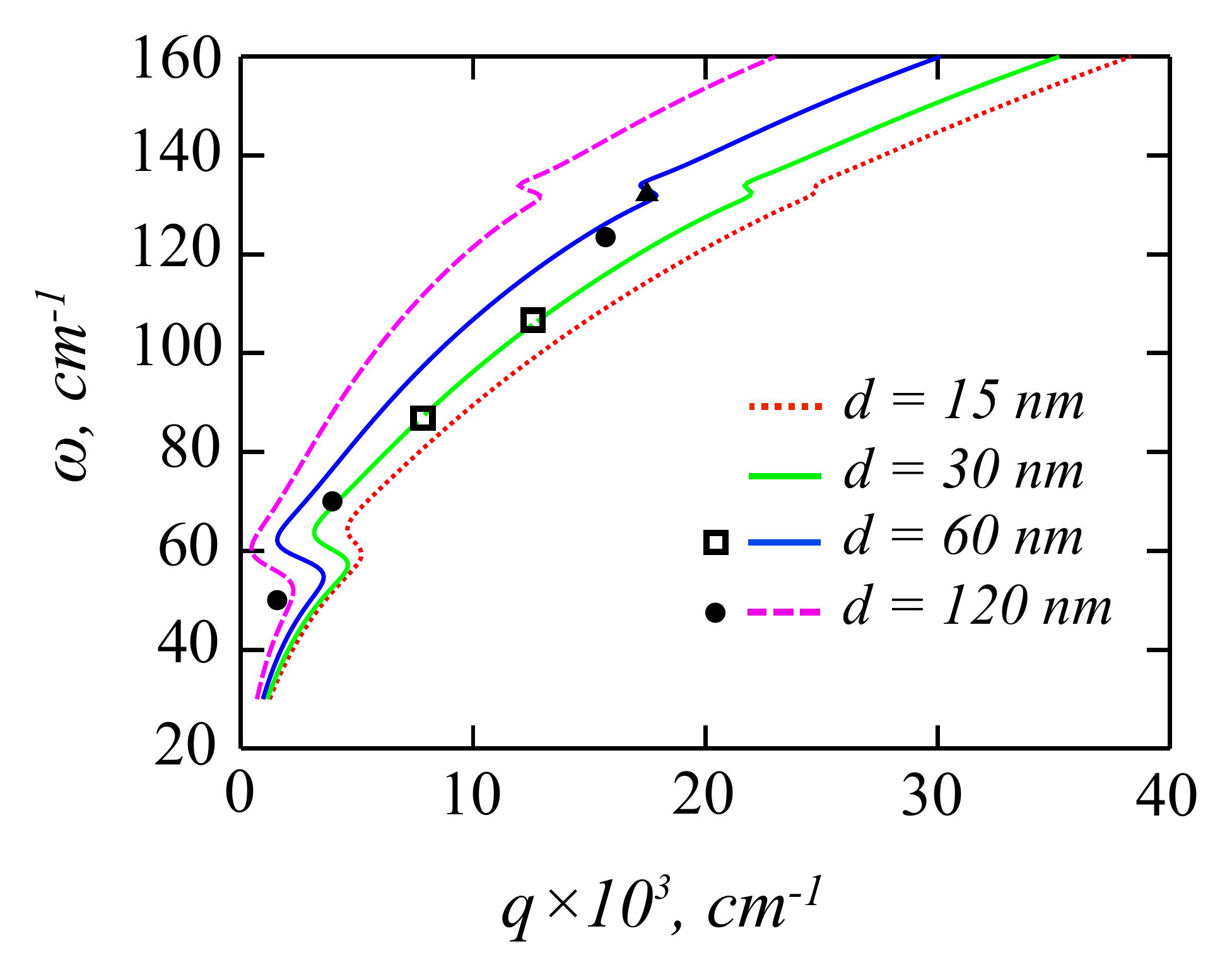}
		\label{fig:subfig4}}
		\qquad
    \caption{Dispersion relations of SPPs in Bi$_2$Se$_3$ films for various combinations of film thickness and the Fermi level. Cirlce, square and triangle marks represent the experimental values from Ref.~\cite{DiPietro}.}
    \label{fig:DispersionBi2Se3}
\end{figure}

\begin{figure}[t]
		\subfloat[Subfigure 1 list of figures text][$k_F = 0.15\times 10^{9}$ m$^{-1}$, $E_F = 50$ meV.]{
		\includegraphics[scale=0.5]{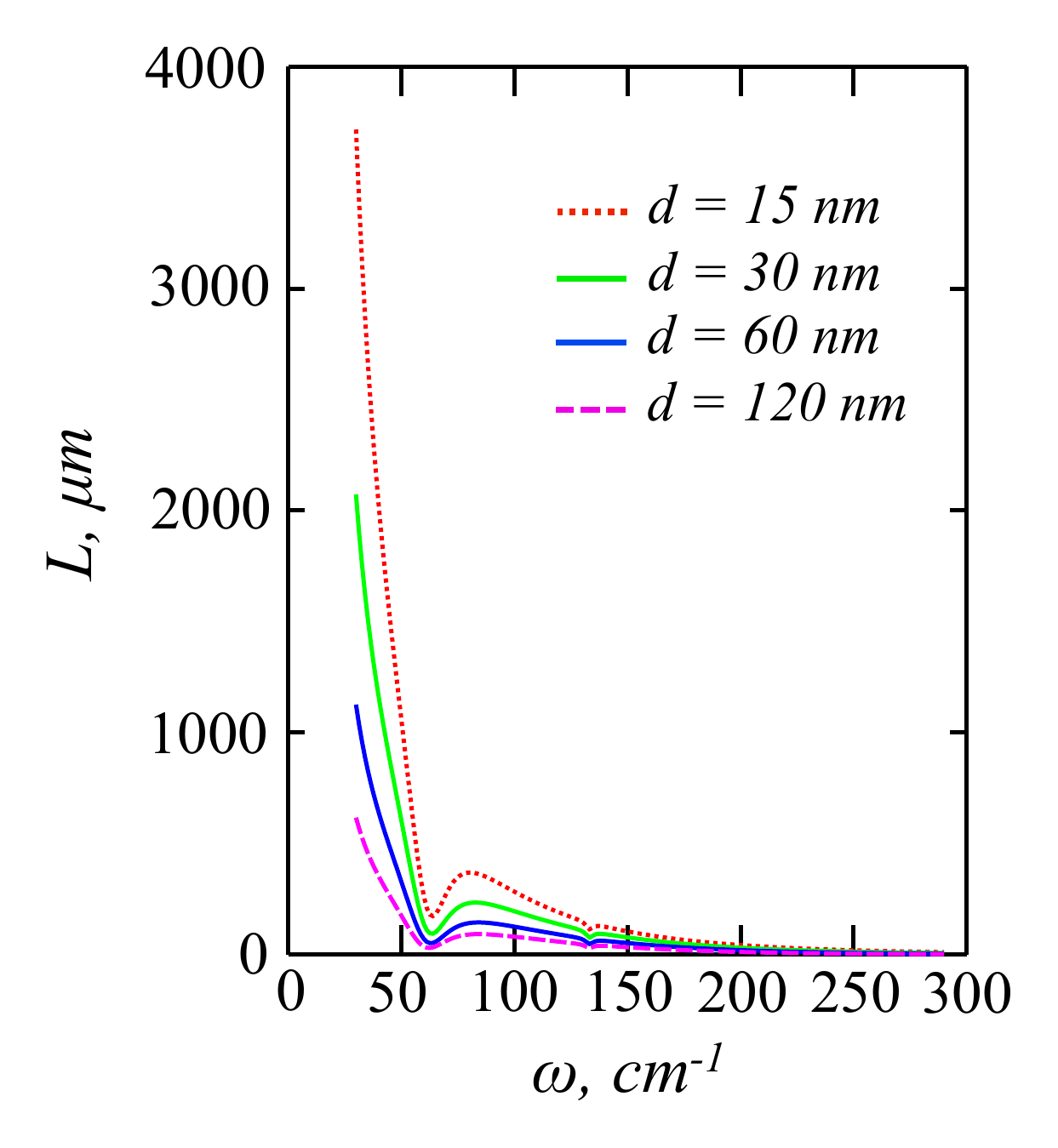}
		\label{fig:subfig1}}
		\qquad
		\subfloat[Subfigure 4 list of figures text][$k_F = 1.52\times 10^{9}$ m$^{-1}$, $E_F = 500$ meV.]{
		\includegraphics[scale=0.5]{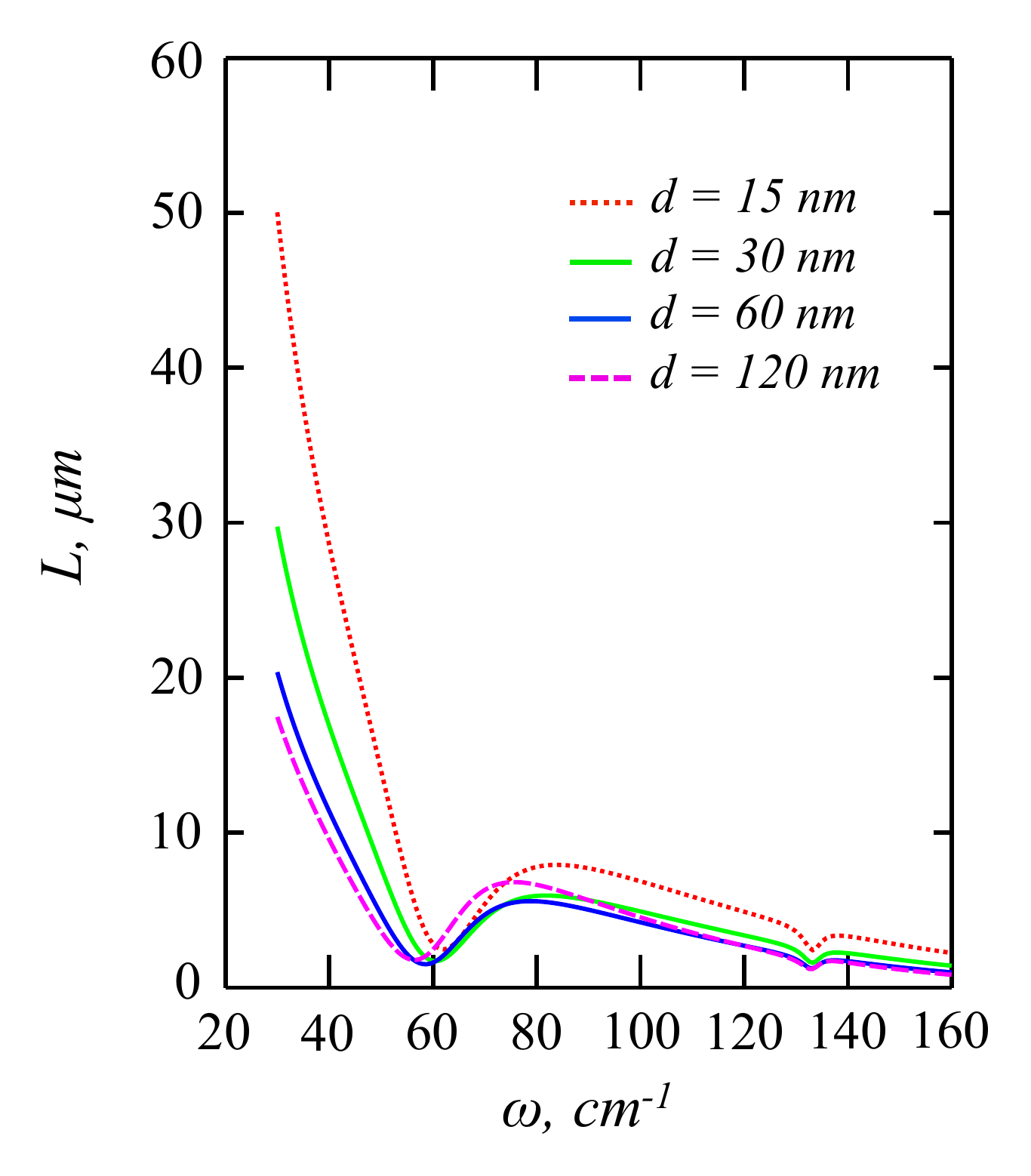}
		\label{fig:subfig4}}
		\qquad
    \caption{Propagation length of SPPs in Bi$_2$Se$_3$ films for various combinations of film thickness and the Fermi level.}
    \label{fig:PropataionBi2Se3}
\end{figure}

In Fig.~\ref{fig:DispersionBi2Se3} dispersion curves $\omega(q)$ are shown for various combinations of TI film thickness, $d$, and the Fermi level, $E_F$, while Fig.~\ref{fig:PropataionBi2Se3} illustrates propagation lengths $L$ for the same parameters. Black dots in Fig.~\ref{fig:DispersionBi2Se3} correspond to the experimental values taken from Ref.~\cite{DiPietro}. While there are visible changes in dispersion curves, the most notable feature is the drastic enhancement of the propagation length as either the film thickness or the Fermi level is decreased.
We note that if the contribution of the Drude term in the model (\ref{eq:DrudeLorentz}) is naively set to zero then similar degrees of sensitivity of dispersion relations $\omega(q)$ and the propagation lengths are exhibited: About fivefold increase of the propagation length happens when there are no free carriers in the bulk, while $\omega(q)$ does not show such a significant change. These results indicate that the propagation length of SPP in Bi$_2$Se$_3$ is very sensitive to the concentration of both surface and bulk carriers. Since the Fermi level can be controlled by gating, this sensitivity suggests a way of tuning the propagation length by almost two orders of magnitude without significantly affecting the SPP dispersion relation.

\begin{figure}
    \centering
		\subfloat[Subfigure 1 list of figures text][Dispersion of SPP in thin films of Bi$_2$Se$_3$ with thicknesses 60 and 120 nm. Cirlce, square and triangle marks correspond to the experimental results \cite{DiPietro}.]{
		\includegraphics[scale=0.5]{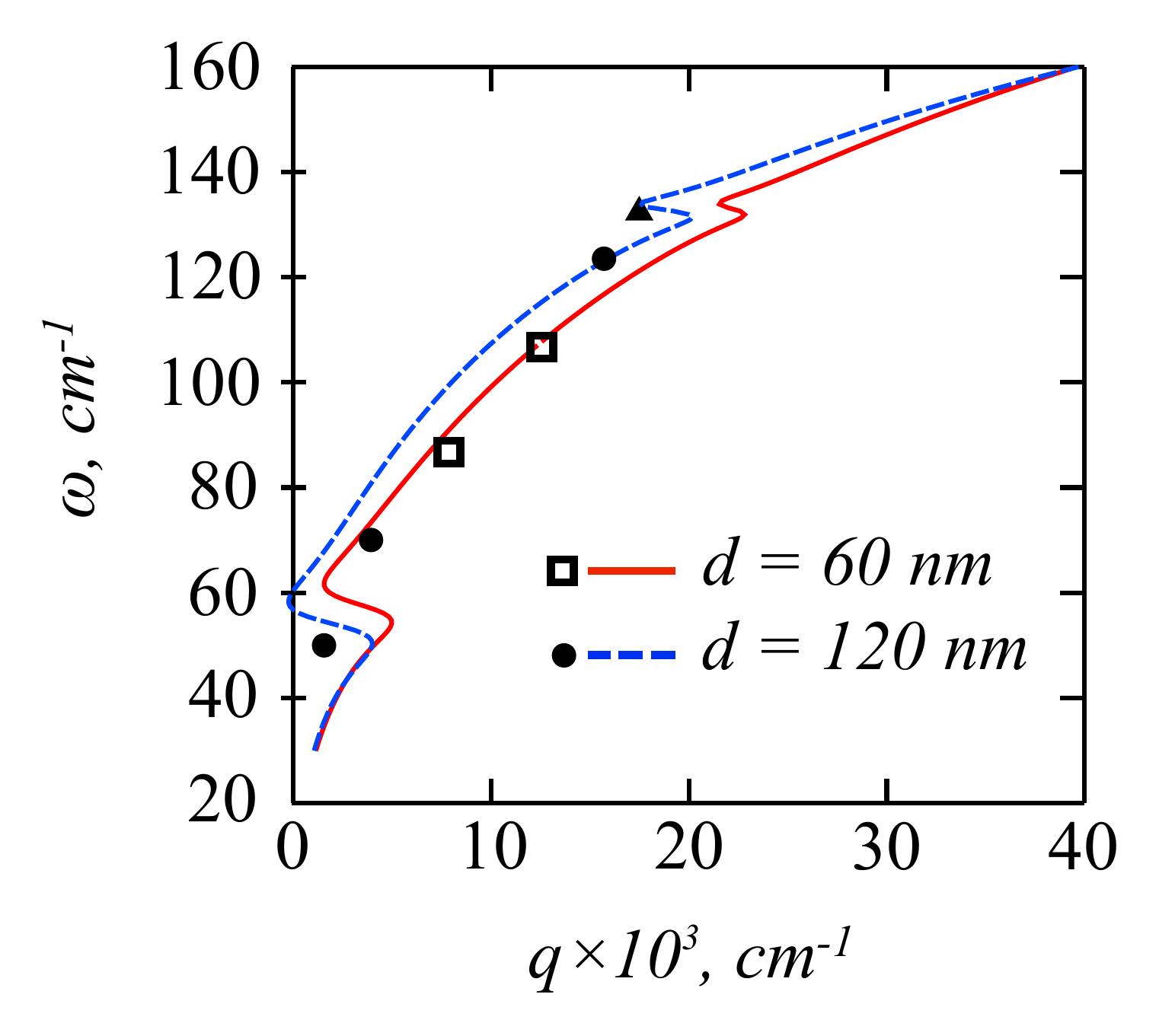}
		\label{fig:subfig1}}
		\qquad
		\subfloat[Subfigure 2 list of figures text][Propagation lengths of SPP in thin film of Bi$_2$Se$_3$ with thicknesses 60 and 120 nm..]{
		\includegraphics[scale=0.5]{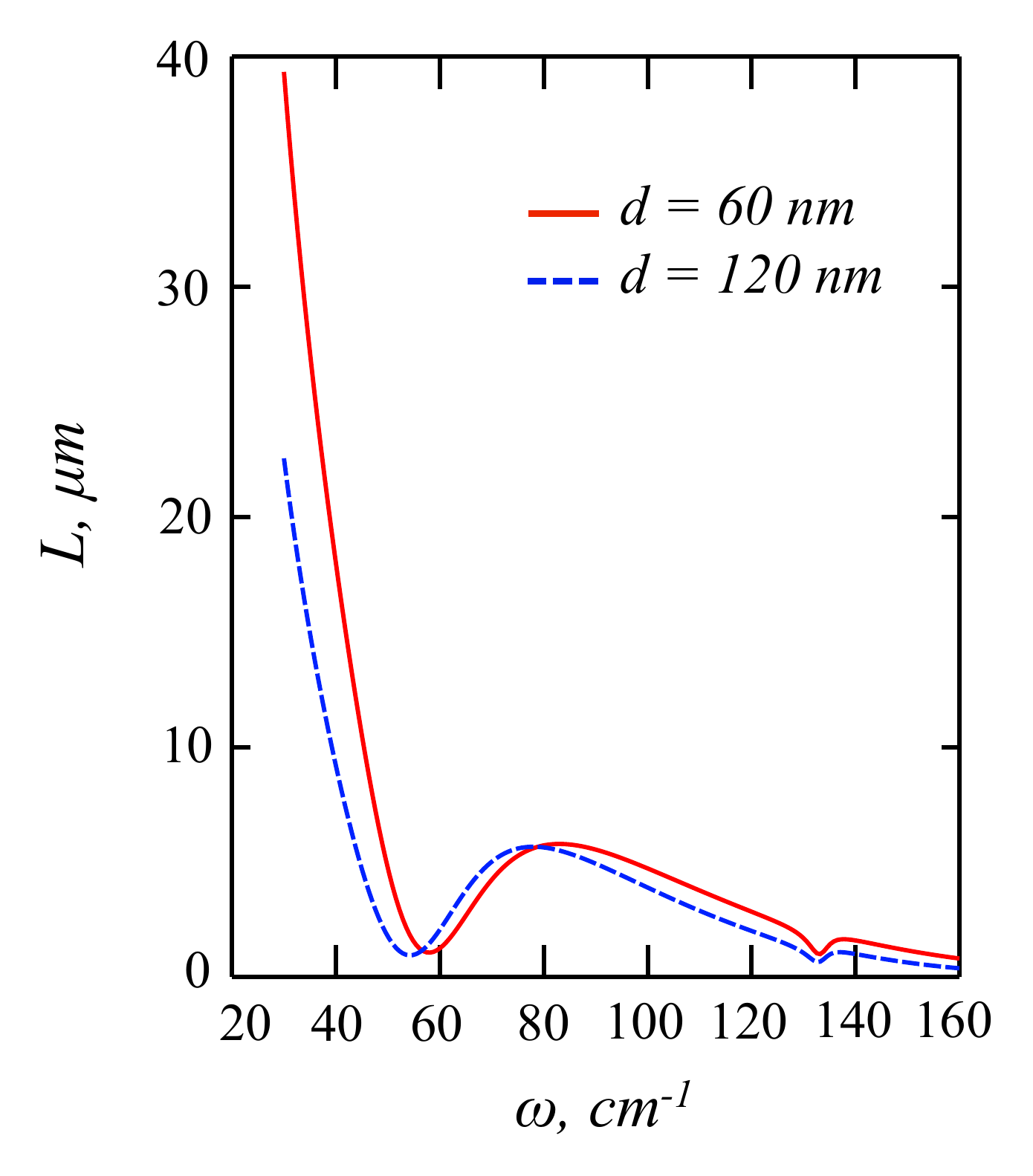}
		\label{fig:subfig2}}
		\qquad		
    \caption{Dispersion curves and propagation lengths of SPPs in Bi$_2$Se$_3$ thin films, calculated using the Drude-Lorentz model and the parameters reported in Ref.~\cite{DiPietro} ($k_F = 1.37\times 10^{9}$ m$^{-1}$, $E_F = 541$ meV).}
    \label{fig:DiPietroBi2Se3}
\end{figure}

We next perform calculations of the dispersion curves and propagation lengths for the experimental parameters given in Ref.~\cite{DiPietro}: $v_F = 6\times 10^5$ m/s, $k_F = 1.37\times 10^9$ m$^{-1}$, $d = 60$ and 120 nm. Relevant parameters for the Drude-Lorentz model are given in Table \ref{tbl:DLDiPietro} and agree well with the numbers given in Ref.~\cite{Wolf} and reproduced in Table \ref{tbl:DLWolf}. The high-frequency Lorentz oscillator with $\Omega_{gap} \approx 2030$ cm$^{-1} = 252$ meV can not be neglected because its contribution to $\epsilon_{TI}$ for the frequencies 30 cm$^{-1}$-200 cm$^{-1}$ is significant ($\epsilon \approx 30$) and does not vary appreciably. We thus take the values for the oscillators $\alpha$ and $\beta$ from Ref.~\cite{DiPietro} (Table \ref{tbl:DLDiPietro}), while borrowing the value for $\Omega_{gap}$ from Ref.~\cite{Wolf} (Table \ref{tbl:DLWolf}). The dispersion curves and propagation lengths are given in Fig.~\ref{fig:DiPietroBi2Se3}. To achieve a reasonable fit to the experimental data for both films thicknesses the magnitude of the Drude term was reduced to 40\% of its reported value. Since the contribution of this term is not precisely known (see the discussion above) this adjustment is justified. This calculation demonstrates that a reasonable agreement of the theory and experiment may be achieved without additional two-dimensional electron gas proposed in Ref.~\cite{Stauber}.

\subsection{Bi$_2$Te$_3$ and Sb$_2$Te$_3$}

Optical characterization of Bi$_2$Te$_3$ has been performed in a number of works \cite{Wolf, Richter, Kaddouri}, with substantial variations of the reported values for optical parameters. Such variations indicate that for realistic calculations the bulk dielectric function must be estimated from reflectance or ellipsometry measurements on a particular sample. In our calculations the hybrid Drude-Lorentz model was used, with parameters from both Richter \cite{Richter} and Wolf \cite{Wolf} (see Tables \ref{tbl:DLRichterBi2Te3} and \ref{tbl:DLWolfBi2Te3}, respectively). This is done to capture contributions from in-plane IR-active vibrational modes present in Bi$_2$Te$_3$ around 50 cm$^{-1}$ and 90 cm$^{-1}$, as well as describe the  plasma edge clearly observed near 500 cm$^{-1}$. 

\begin{table}
\small
\begin{center}
    \begin{tabular}{@{} llll}
    \hline\hline
    Oscillator & $\omega_0$, cm$^{-1}$ & $\omega_p$, cm$^{-1}$ & $\gamma$, cm$^{-1}$ \\ \hline
    Drude & NA & NA & NA \\
    Lorentz, $\alpha$ & 50 & 716 & 10 \\
    Lorentz, $\beta$ & 95 & 116 & 15 \\
	Lorentz, $\Omega$ & NA & NA & NA \\
		\hline\hline
    \end{tabular}	
	\caption{Parameters of the Drude-Lorentz model with $\epsilon_\infty = 85$, extracted from the reflectance measurements on Bi$_2$Te$_3$ at room temperatures in far-IR \cite{Richter}.}
	\label{tbl:DLRichterBi2Te3}
\end{center}
\end{table}

\begin{table}
\small
\begin{center}
    \begin{tabular}{@{} llll}    
    \hline\hline
    Oscillator & $\omega_0$, cm$^{-1}$ & $\omega_p$, cm$^{-1}$ & $\gamma$, cm$^{-1}$ \\ 
    \hline
    Drude & 0 & 5651.5 & 111.86 \\
    Lorentz, $\alpha$ & NA & NA & NA \\
    Lorentz, $\beta$ & NA & NA & NA \\
	Lorentz, $\Omega$ & 8386.6 & 66024 & 10260 \\
		\hline\hline
    \end{tabular}	
	\caption{Parameters for the Drude-Lorentz model with $\epsilon_\infty = 1$ extracted from the reflectance measurements on Bi$_2$Te$_3$ at room temperatures in far- and mid-IR \cite{Wolf}.}
	\label{tbl:DLWolfBi2Te3}
\end{center}
\end{table}

\begin{figure}
    \centering
		\subfloat[Subfigure 1 list of figures text][$k_F = 0.15\times 10^{9}$ m$^{-1}$, $E_F = 50$ meV.]{
		\includegraphics[scale=0.4]{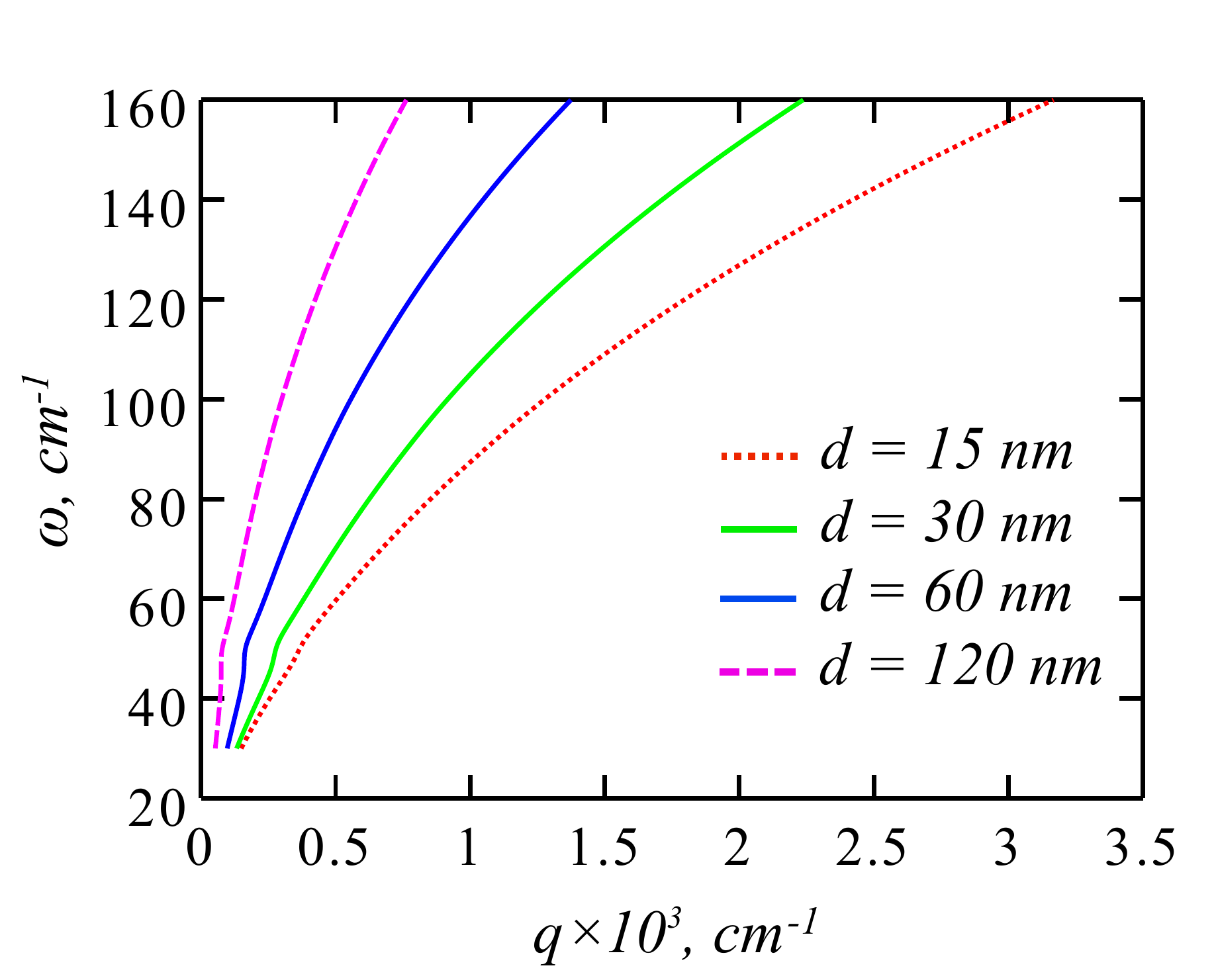}
		\label{fig:subfig1}}
		\qquad
		\subfloat[Subfigure 4 list of figures text][$k_F = 1.52\times 10^{9}$ m$^{-1}$, $E_F = 50$ meV.]{
		\includegraphics[scale=0.4]{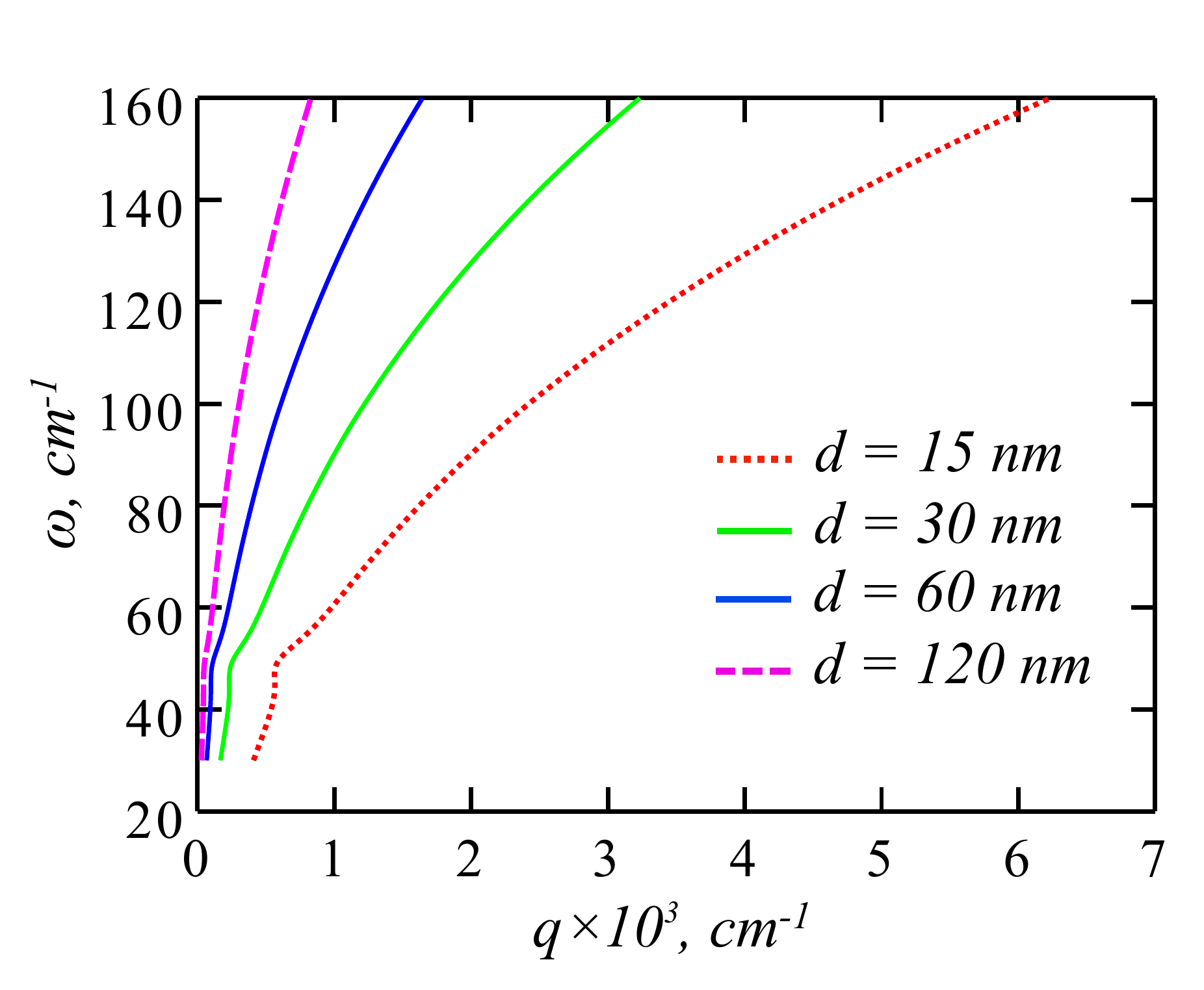}
		\label{fig:subfig4}}
		\qquad
    \caption{Dispersion relations of SPPs in Bi$_2$Te$_3$ films for various combinations of film thickness and the Fermi level. }
    \label{fig:DispersionBi2Te3}
\end{figure}
\begin{figure}
    \centering
		\subfloat[Subfigure 1 list of figures text][$k_F = 0.15\times 10^{9}$ m$^{-1}$, $E_F = 50$ meV.]{
		\includegraphics[scale=0.55]{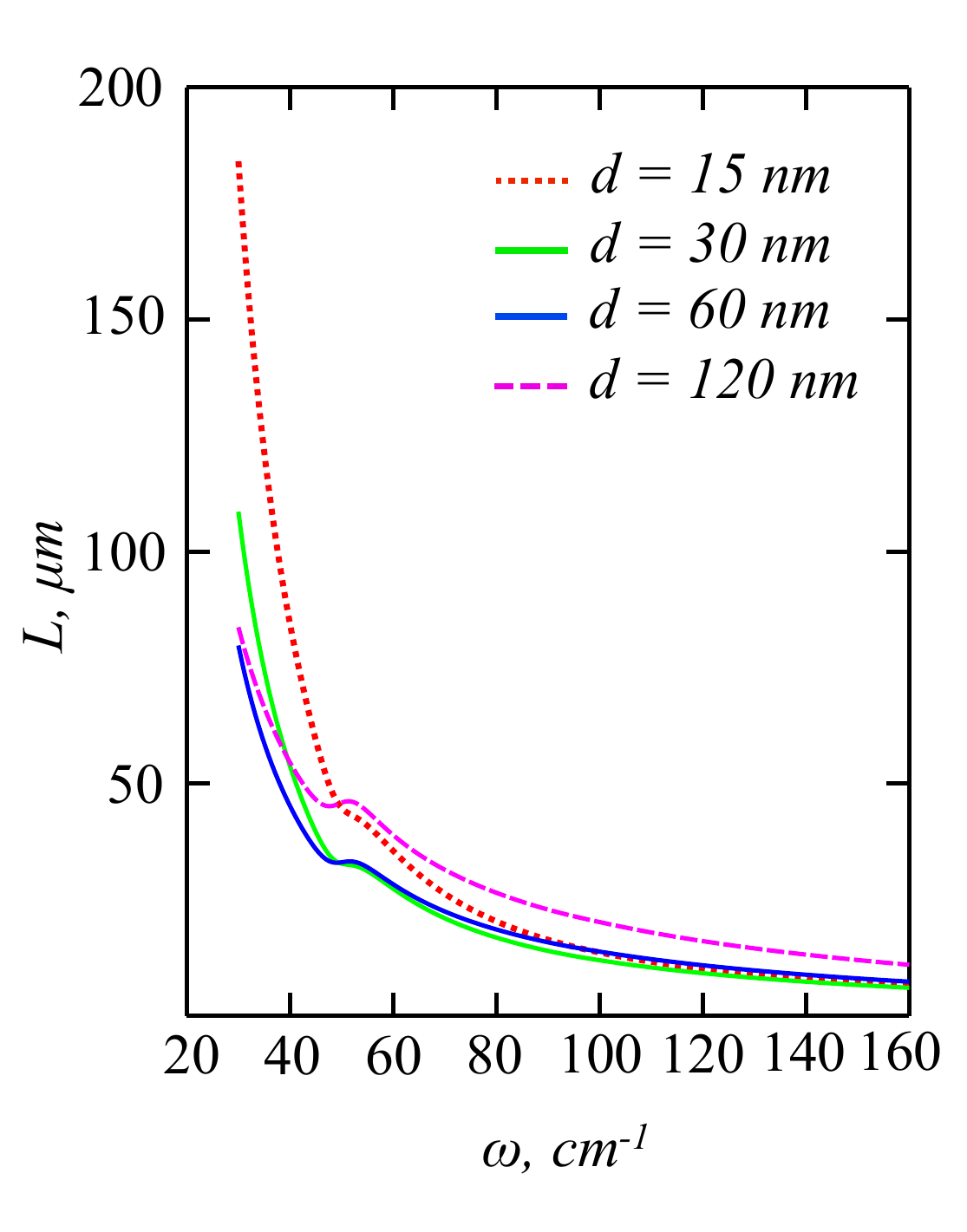}
		\label{fig:subfig1}}
		\qquad
		\subfloat[Subfigure 4 list of figures text][$k_F = 1.52\times 10^{9}$ m$^{-1}$, $E_F = 50$ meV.]{
		\includegraphics[scale=0.4]{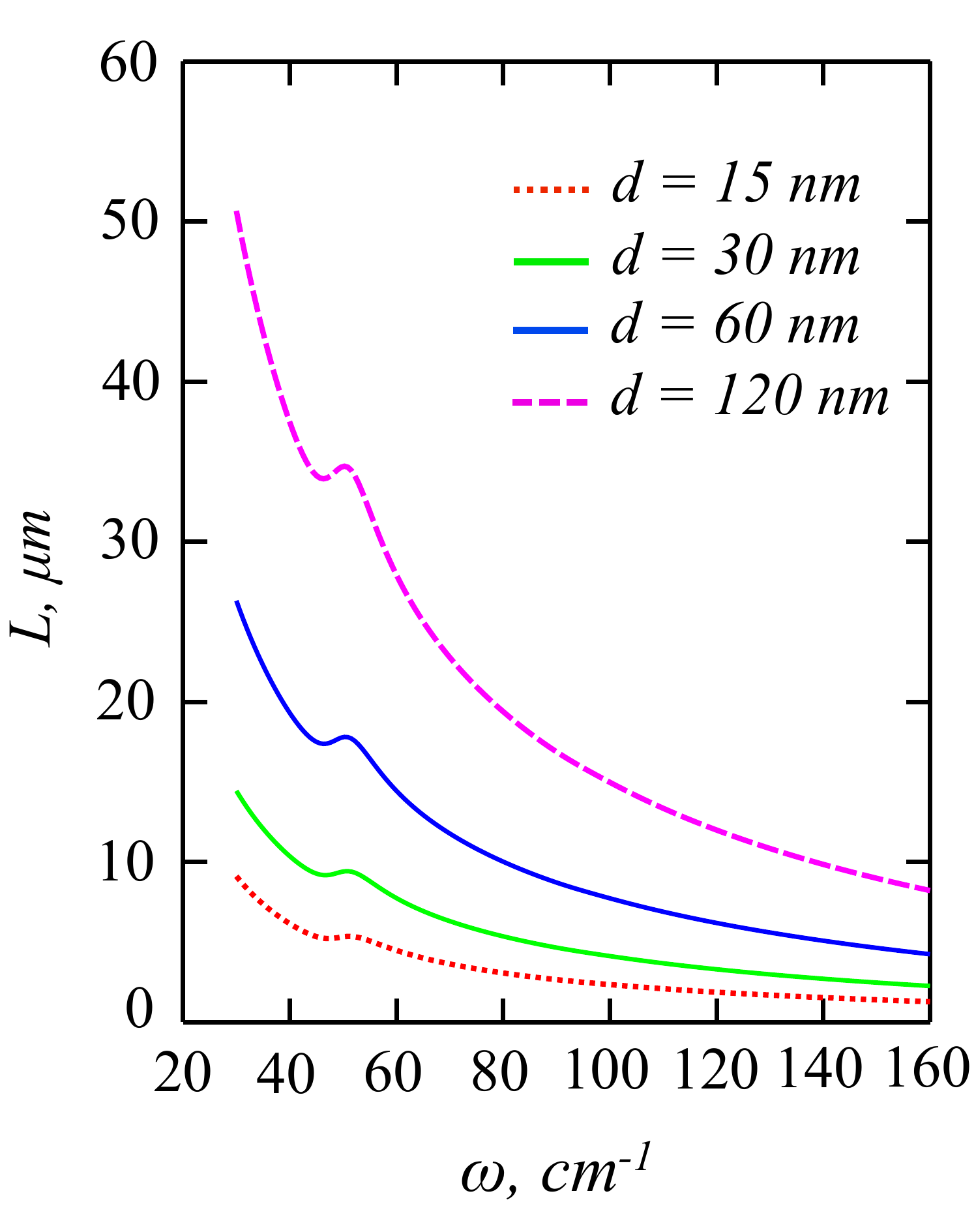}
		\label{fig:subfig4}}
		\qquad
    \caption{Propagation lengths of SPPs in Bi$_2$Te$_3$ films for various combinations of film thickness and the Fermi level.}
    \label{fig:PropataionBi2Te3}
\end{figure}

The results of calculations are presented in Fig.~\ref{fig:DispersionBi2Te3} and Fig.~\ref{fig:PropataionBi2Te3}. Compared to Bi$_2$Se$_3$, Bi$_2$Te$_3$ demonstrates 1) greater sensitivity of dispersion curves $\omega(q)$ to the film thickness; and 2) significantly smaller propagation lengths for almost all values of the Fermi level. These differences become less pronounced if the Drude term is omitted, suggesting that overall higher level of free electrons in the bulk of Bi$_2$Te$_3$  strongly affects the propagation of the SPP in this material.

\begin{table}
\begin{center}
    \begin{tabular}{@{} llll}
    \hline\hline
    Oscillator & $\omega_0$, cm$^{-1}$ & $\omega_p$, cm$^{-1}$ & $\gamma$, cm$^{-1}$ \\ \hline
    Drude & 0 & 6906.7 & 183.69 \\
    Lorentz, $\alpha$ & 67.03 & 1498.0 & 10.0 \\
    Lorentz, $\beta$ & NA & NA & NA \\
	Lorentz, $\Omega_{gap}$ & NA & NA & NA \\
		\hline\hline
    \end{tabular}	
	\caption{Parameters for the Drude-Lorentz model with $\epsilon_\infty = 51$ extracted from the reflectance measurements on Sb$_2$Te$_3$ at room temperatures in far and mid-IR \cite{Wolf, Richter}.}
	\label{tbl:DLWolfRichterSb2Te3}
\end{center}
\end{table}

Similar to Bi$_2$Te$_3$ the data on Sb$_2$Te$_3$ \cite{Wolf, Richter} was combined to make the hybrid Drude-Lorentz model with the parameters presented in Table~\ref{tbl:DLWolfRichterSb2Te3}. The behavior of the dispersion curves and propagation lengths is expectedly similar to Bi$_2$Te$_3$, given the similarity between the optical parameters for these materials. We therefore do not include the results of calculations here.

\subsection{SPP in Superlattice}
\begin{figure}[!h]
    \centering	
	 \includegraphics[scale=0.25]{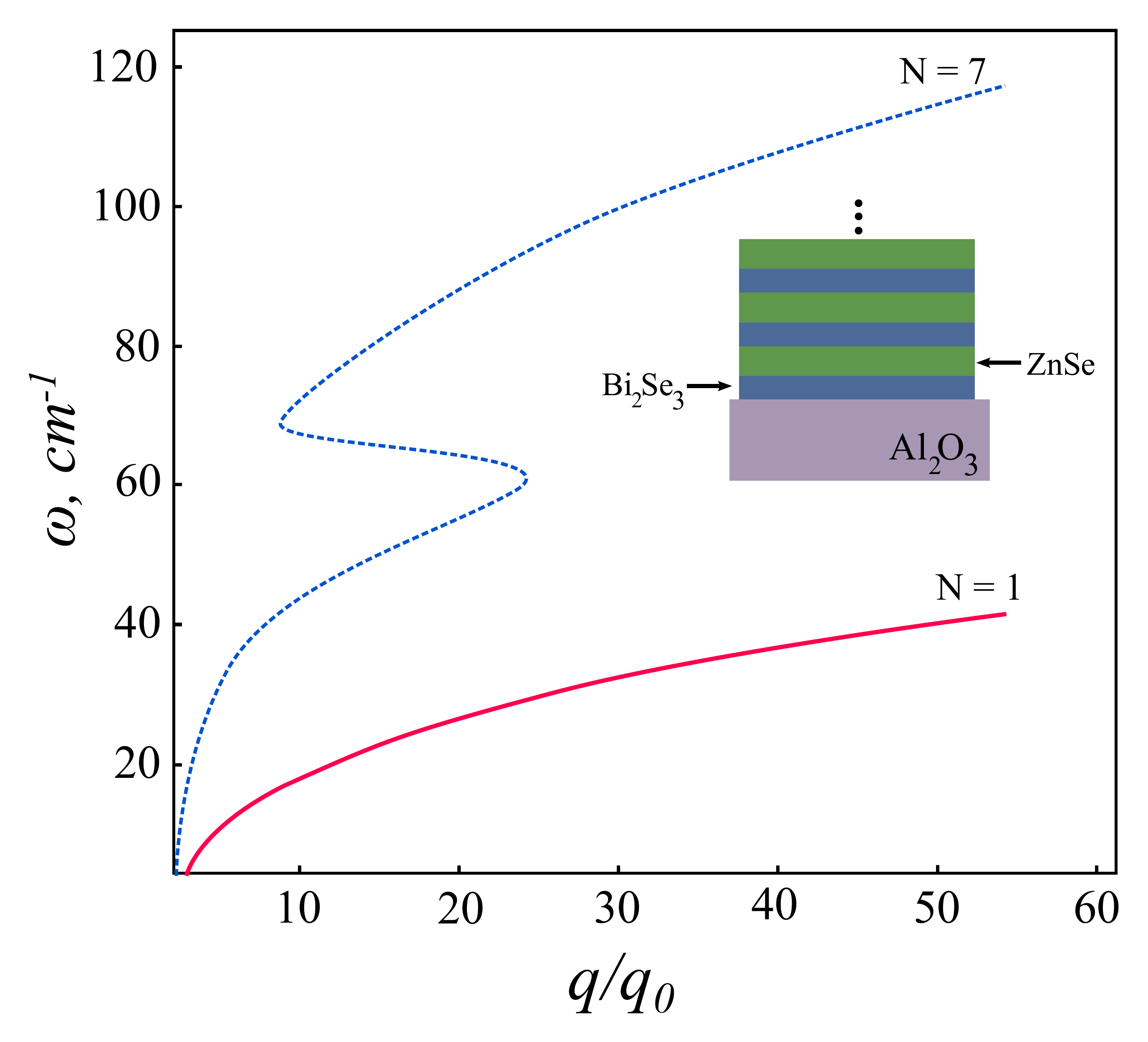}
	 \caption{Tuning SPP dispersion by varying the number of units cells in superlattice of TI and dielectric (ZnSe in this case). Transfer matrix calculations for the structure reported in Ref.~\cite{Chen}.}
    \label{fig:superLattice}
\end{figure}

The study of SPPs in multi-layered graphene waveguides \cite{Smirnova} shows that in the limit of long wavelengths ($qd\ll 1$) the effective optical conductance increases linearly with the number of layers in the stack. Therefore it seems plausible to use superlattices made of alternating layers of thin films of TI and dielectric in order to tune the dispersion curve of a fundamental mode of SPP by growing the required number of layers. The growth of such superlattice using Bi$_2$Se$_3$ films has been reported by Chen et al. \cite{Chen}. Since the SPP frequency grows with conductance, increasing the number of unit cells in the superlattice allows to shift-up the SPP frequency for a given wave-vector $q$. The illustration of such a shift is given in Fig.~\ref{fig:superLattice}. The energy dispersion curves for SPP in layers of Bi$_2$Se$_3$/ZnSe (9 nm and 10 nm thick, respectively) were calculated using transfer matrix method. Optical conductance of TI surface was modeled using the Drude-like expression $G(\omega) = i\sigma_0/\bigl\lbrack 1+(4\hbar\omega\sigma_0/\mu)^2\bigr\rbrack$ with $\sigma_0 = 138 G_0$, $G_0 = 2e^2/h$, and $\mu = 0.500$ eV. Bulk dielectric function of ZnSe in far-IR is taken from Ref.~\cite{ZnSeFarIR}. The results of calculations support the idea of changing the SPP dispersion relations by increasing number of ``unit cells'' in the superlattice.

\section{\label{sec:conclusions}Conclusions}
The results of this work can be summarized as follows:

\begin{enumerate}[i]
	\item Dispersion relations and propagation lengths of SPPs in thin films of Bi$_2$Se$_3$, Bi$_2$Te$_3$, and Sb$_2$Te$_3$ were determined using realistic material parameters in far-IR. Bi$_2$Se$_3$ is identified as the material of choice if larger propagations lengths are desired.	
	\item Key parameters influencing the propagation length are found to be 1) the Fermi level, $E_F$, and 2) film thickness. Lowering $E_F$ by gating is a feasible way to control the propagation lengths of SPPs. Additional enhancement of propagation lengths can be achieved when working with thin films ($d\le$ 20 nm). As an example, SPP in 15 nm thin film of Bi$_2$Se$_3$ with $E_F$ = 50 meV will propagate about 100 times farther than in 120 nm film with $E_F$ = 500 meV.	
	\item The disagreement between the theory of SPP in TIs \cite{Stauber} and the first experimental measurements on Bi$_2$Se$_3$ \cite{DiPietro} is removed by simply considering realistic optical properties of Bi$_2$Se$_3$. We stress again the importance of allowing the bulk dielectric function to vary when analyzing  experiments or discussing possible applications.	
	\item It was demonstrated that stacking of TI films and dielectrics into  superlattice is a promising way to modify dispersion relations of SPPs. This raises an intriguing question: Can stacking (possibly combined with other controlling factors) shift the SPP frequency up into the spectral range of e.g. quantum cascade lasers. Such a scenario, if experimentally realized, may open a new venue for manipulating SPPs in terahertz range.
\end{enumerate}

\begin{acknowledgments}
We would like to thank Dr.~P.~Ghaemi and Dr.~A.~Punnoose for helpful discussions. We wish to acknowledge funding from CUNY ASRC Joint Seed Program. Research was carried out in part at the Center for Functional Nanomaterials, Brookhaven National Laboratory, supported by the U.S. Department of Energy, Office of Basic Energy Sciences, under Contract No. DE-SC0012704. This work was also supported in part by NSF DMR-1420634 and DOD-W911NF-13-1-0159. Vinod Menon would like to acknowledge funding from NSF through the EFRI-2DARE program (EFMA-1542863).
\end{acknowledgments}

\bibliography{mainSPPV4}

\end{document}